\documentclass{aa} 
\usepackage{graphicx}
\usepackage{txfonts}
\usepackage{natbib}

\begin{document}
   \title{Non-perturbative effect of rotation \\on dipolar mixed modes in red giant stars}
   \author{R-M. Ouazzani
          \inst{1, 2},
          M.J. Goupil\inst{2},
          M-A. Dupret\inst{1}
          \and
          J.P. Marques\inst{3,2},
          }
   \institute{Institut d'Astrophysique et de G\'eophysique de l'Universit\'e de Li\`ege, All\'ee du 6 Ao\^ut 17, 4000 Li\`ege, Belgium
         \and
             Observatoire de Paris, LESIA, CNRS UMR 8109, F-92195 Meudon, France
         \and
             Georg-August-Universit\"at G\"ottingen, Institut f\"ur Astrophysik, Friedrich-Hund-Platz 1, D-37077 G\"ottingen, Germany\\
              \email{rhita-maria.ouazzani@obspm.fr}
         }
   \date{Received , 2012; accepted , 2013}

  \abstract
   {The space missions CoRoT and Kepler provide high quality data  that allow us to test the transport of angular momentum in stars by the seismic determination of the internal rotation profile. } 
   {Our aim is to  test the validity of the seismic diagnostics for red giant rotation  that are  based on a perturbative method and to investigate the oscillation spectra when the validity does not hold.}
   {We use a non-perturbative approach implemented in the ACOR code (Ouazzani et al. 2012) 
   that accounts for the effect of rotation on pulsations, and solves the pulsation eigenproblem 
   directly for dipolar oscillation modes.}
   {We find that the limit of the perturbation to first order can be expressed in terms of the rotational splitting compared to the frequency separation between consecutive dipolar modes. Above this limit, non-perturbative computations are necessary but only one term in the spectral expansion of modes is sufficient as long as the core rotation rate remains significantly smaller than the pulsation frequencies. Each family of modes with different azimuthal symmetry, $m$, has to be considered separately. In particular, in case of rapid core rotation, the density of the spectrum differs significantly from one $m$-family  of modes to another, so that the differences between the period spacings associated with each $m$-family can constitute a promising guideline toward a proper seismic diagnostic for rotation.}
  {}
 \maketitle
%

\section{Introduction}

Seismic measurements of rotation profiles inside the Sun as well as stars provide tight constraints on models of  transport of angular momentum \citep[][and references therein]{Pinsonneault1989,Zahn1992,Zahn1997,Talon2008}. 
In particular, stars in late stages of evolution, due to the highly condensed core, oscillate with non-radial modes that have a mixed character: they behave as p modes in the envelope and as g modes in the core. These modes, also known as {\it mixed modes}, are of particular interest for the determination of the rotation profile throughout the star, as they carry the signature of the star's innermost layers and are detectable at the surface. The CoRoT \citep{Baglin2006} and \emph{Kepler} \citep{Borucki2010} spacecrafts have dramatically improved the quality of the available asteroseismic data. Several recent studies reported the detection of mixed modes that are split by rotation in a subgiant \citep{Deheuvels2012} and in several red giants \citep{Beck2012,Mosser2012a,Mosser2012b} observed with \emph{Kepler}. A large number of these stars exhibit frequency spectra that show a quite simple structure where symmetric patterns around axisymmetric modes are easily identified. They are interpreted as multiplets of modes split by rotation. The rotational splittings, i.e. the frequency spacing related to the lift of degeneracy caused by rotation is then used to determine the core rotation. The values of the corresponding splittings are quite small and the use of the lowest  order approximation to derive the splittings from stellar models can be justified. Such studies led to the determination of unexpectedly low central rotation frequencies (of few hundreds of nHz). These results are in strong disagreement with the core rotation frequencies predicted by evolutionary models, which are of the order of few dozens of $\rm \mu$Hz \citep[][]{Eggenberger2012, Marques2013}. They show that the transport processes currently included in stellar models are not able to spin down the core of red giant stars enough to explain the slowly core rotating red giants. On the other hand, a large set of red giant stars show complex frequency spectra \citep{Mosser2012b}, in particular with non symmetric multiplets and therefore are likely rotating fast. Their rotation must then be investigated  with non-perturbative methods.

In this context, we first report  on the relevance of using a first order approach  for the inference of rotation from seismic spectra of red giant stars with slowly to rapidly rotating core. When not relevant, we adopt the non-perturbative approach in order to shed light on the behaviour of splitted mixed modes in red giants spectra.

\section{Theoretical frequency spectra for red giants}

\begin{figure}[t!]
  	    {\hspace*{-1.3cm}\includegraphics[scale=0.42, angle=-90]{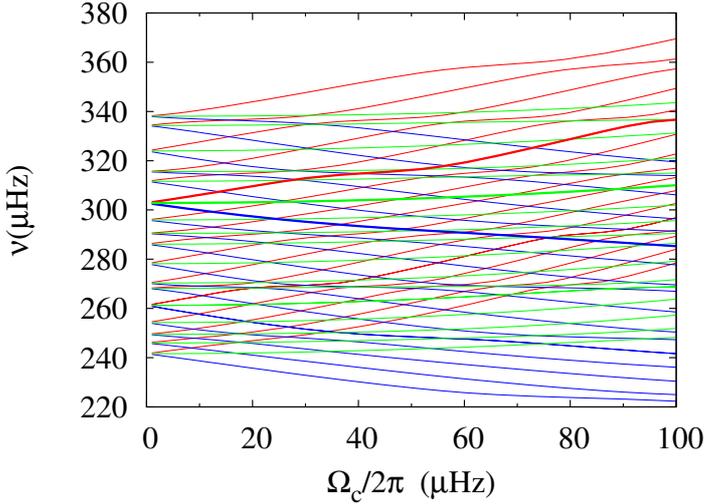}}
            \caption{\label{Fig_suivi_rot}
              Pulsation frequencies in the inertial frame $\nu$ (in $\mu$Hz) versus core rotation frequency $\Omega_c / 2 \pi$ (in $\mu$Hz) for dipolar multiplets. Red curves refer to $\rm m=-1$ (prograde) modes, green curves to $\rm m=0$ modes (axis-symmetric) , and blue curves to $\rm m = 1$ (retrograde) modes, for central rotation frequency ranging from 0 to 100 $\mu$Hz.}
\vspace*{-0.5cm}
\end{figure}

This work is based on the study of a model, (model M$_1$with mass 1.3 M$_{\odot}$ and radius 3.6 R$_{\odot}$), at the bottom of the red giant branch.  The stellar model is computed with the CESTAM code \citep[Code d'Evolution Stellaire, avec Transport, Adaptatif et Modulaire, ][]{Marques2013}. Transport of angular momentum induced by rotation is included according to \cite{Zahn1992}. The central rotation rate in this model is ${\rm \Omega_c / 2 \pi \simeq 180 \mu}$Hz, while the surface convective region rotates at a rate of $1\, \mu$Hz. The distortion of the model due to the centrifugal force  can be neglected everywhere in red giant stars.  For the model M$_1$, scaling laws \citep{Kjeldsen1995} give a frequency of maximum power around $\rm \nu_{max} = 289\, \rm \mu$Hz and a large separation of $\Delta \nu = 23 \,\rm \mu$Hz. We then compute frequencies ranging between $\rm \nu_{max} \pm 2 \Delta \nu$. In this frequency range, the impact of the Coriolis force remains small except in the very inner layers of the star where it can significantly affect the modes. In order to investigate the effect of core rotation on the frequency spectrum, we compute sets of frequencies for model M$_1$ for a sequence of rotation profiles. This sequence is obtained by dividing the whole rotation profile given by CESTAM for model M$_1$ by constant factors. The oscillation frequencies are calculated by the non-perturbative pulsation code ACOR \citep[Adiabatic Code of Oscillations including Rotation, ][]{Ouazzani2012b}. The eigenmodes are obtained as a result of the coupling of spherical harmonics. In what follows, M coupling terms means expansion on the $\ell=1,3, ...,2 \rm M+1$ spherical harmonics for the scalars and the poloidal velocity component and $\ell=2,4,...,2 \rm M+2$ for the toroidal velocity component.  

Figure \ref{Fig_suivi_rot} shows the frequencies of several dipolar ($\ell=1$) multiplets with increasing rotation frequency, computed using one coupling term ($M=1$) in the spectral expansion. Starting at low rotation rate, due to the combined action of the Doppler effect and the Coriolis force, in the inertial frame, prograde ($m=-1$) modes are shifted towards higher frequencies, whereas retrograde ($m=+1$) modes are shifted towards lower frequencies. Crossings between modes of different symmetry occur approximately at $\Omega_c/2\pi \gtrsim 8 \mu$Hz (for model M$_1$). As a result, for core rotation frequencies above $8 \mu$Hz, modes of different $m$ are no longer gathered by {\it original} multiplets, i.e. triplets of modes which have the same degenerate frequency without rotation.

\begin{figure}[t!]
{\includegraphics[scale=0.8, angle=-0]{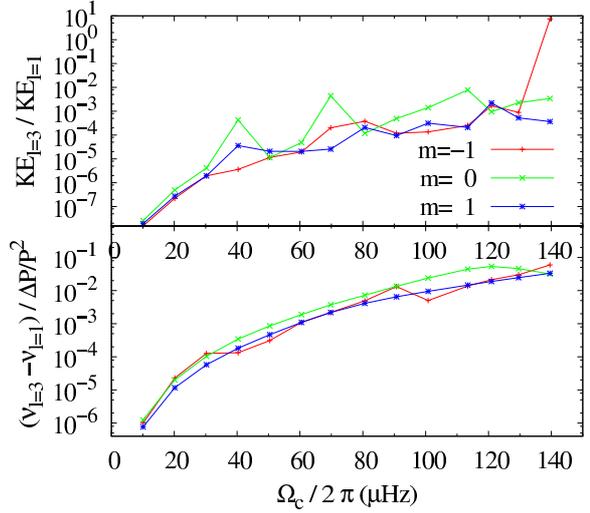}
} 
  \caption{{\it Top}: Ratio between the contributions to kinetic energy of the $\ell=3$ component and the $\ell=1$ component of a dominated $\ell=1$ mode. {\it Bottom}: Difference between the pulsation frequencies computed with three coupling terms ($M=3$) and with one coupling term ($M=1$), divided by the frequency separation between two consecutive mixed modes. Computed for the triplet indicated in thick lines in Fig.~(\ref{Fig_suivi_rot}).}
\label{Fig_M1-M3}
\end{figure}

The choice of $M=1$ coupling term in the spectral expansion is indeed  sufficient for most of the rotation profiles investigated  here. This is illustrated in Fig. \ref{Fig_M1-M3} which shows the impact of the number of spherical harmonics used in the eigenmodes expansion. Fig. \ref{Fig_M1-M3} present the  results of non-perturbative calculations using three coupling terms $(M=3$). Most modes are clearly dominated by their $\ell=1$ component (Fig. \ref{Fig_M1-M3}, top). Figure \ref{Fig_M1-M3} (bottom) shows that the frequency difference between calculations including one and three spherical harmonics are smaller then the frequency difference between two consecutive mixed modes (for $\Omega_c / 2 \pi<100 \mu$Hz by two orders of magnitude).  From now on, we adopt $M=1$.

The trapping of modes essentially depends on their frequency \citep[see for instance][]{Unno1989}. As seen in Fig.~\ref{Fig_suivi_rot}, when $\Omega_c$ increases, so does the frequency difference between the three members of a triplet and the trapping of these members can be significantly different, to such an extent that they end up with very different p-g nature. This is illustrated in Fig. \ref{Ec_trip}, where the kinetic energy corresponding to modes of different $m$ around $\nu_{max}$ is plotted. It gives an indication of the p-g nature of modes: the p-dominated modes (referred as p-m modes) correspond to the minima of energy, while the g-dominated ones (g-m modes) are associated with the maxima. The three modes circled in black belong to the same {\it original} multiplet, and show different p-g nature. The prograde and the retrograde modes are g-m modes, while the $m=0$ one is a p-m mode. This change of nature induced by rotation, which depends on the azimuthal order, can also be characterized by the number of nodes in the p-mode cavity ($\rm n_p$) and in the g-mode cavity ($\rm n_g$) \citep[calculated according to the Cowling approximation,][]{Cowling1941}. As shown in Fig. \ref{Ec_trip}, the $\rm n_p$ and $\rm n_g$ values of the members of an {\it original} triplet are modified differently. These gains (for $m=0,1$ modes) or losses (for $m=-1$ modes) of nodes occur during gravito-acoustic avoided crossings. When increasing the rotation rate, if the frequencies of two modes of same symmetry become very close, they avoid to cross each other and exchange nature. These avoided crossings need to be taken into account all the more because they mainly affect the p-m modes that are the most likely to be observed. 
Therefore, modes of different azimuthal order $m$ probe differently the stellar interior. Even if they belong to the same triplet with radial order $n$ and degree $\ell$, they can be of very different nature, and using the rotational splitting $\delta \omega_{n,\ell}=(\omega_{n,\ell,m}- \omega_{n,\ell,-m})/2m$ to determine the rotation rate is therefore questionable for red giants.

\section{Slow to moderate core rotating red giants}
\label{S2}

For models with $\Omega_c/2\pi \le $ 8 $\rm \mu$Hz (see Fig. \ref{Fig_suivi_rot}), the frequencies behave linearly with respect to $\Omega_c$. In this range, the sectorial modes ($\ell = \, \mid m \mid$) are symmetrically distributed around the axisymmetric modes, and have the same radial order. 
\begin{figure}[t!]
  \resizebox{\hsize}{!}
{\includegraphics[scale=0.3, angle=-90]{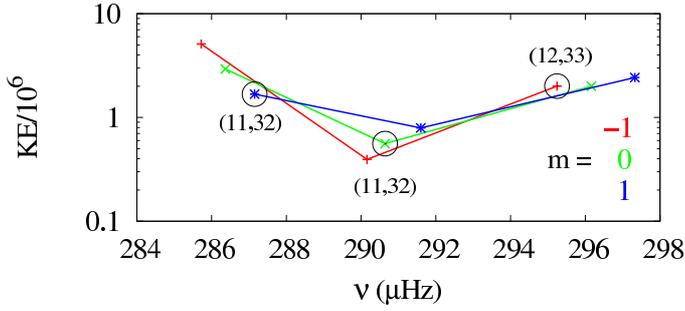}
}  
  \caption{\label{Ec_trip}
For a central rotation frequency of 20$\rm \mu$Hz, kinetic energy of modes around a p-m mode as a function of the mode frequency. The three modes circled in black belong to the same {\it original} triplet. Inside the parentheses are indicated their number of nodes in the p cavity and in the g cavity.}
\vspace*{-0.5cm}
\end{figure}
In this case, a first order perturbation approach approximates well the effects of rotation on the mode frequencies. The pulsation $\omega_{n,\ell,m}$ is then given by $\omega_{n,\ell,m} \, = \, \omega_{n,\ell,0} \, + \, m \delta \omega_{n,\ell}^1$ (\citealt{Cowling1949} and \citealt{Ledoux1949}) where the first order rotational splitting $\delta \omega_{n,\ell}^1$ is expressed as a weighted measure of the star rotation rate: $\delta \omega_{n,\ell}^1 \, = \, \int_0^R \, K_{n,\ell}(r) \, \Omega(r)\, \rm dr$, where $K_{n,\ell}(r)$ are the rotational kernels of the modes. They depend on the equilibrium structure and on the eigenfunctions of the unperturbed modes \citep[see e.g. ][and references therein]{Goupil2009}. This formulation relies on two main assumptions. First, only the Coriolis force is significant and it is enough to account for its contribution to order $O(\Omega/\omega_{n,\ell})$. Second, it is derived from a variational principle \citep[see e.g. ][]{Lynden-Bell1967}, which requires that the eigenfunction of the mode perturbed by rotation is close to the eigenfunction of the corresponding unperturbed mode. This is the case for model M$_1$  for $\Omega_c / 2 \pi < 8 ~\mu$Hz.

\begin{figure}[t!]
{\includegraphics[scale=0.8, angle=-0]{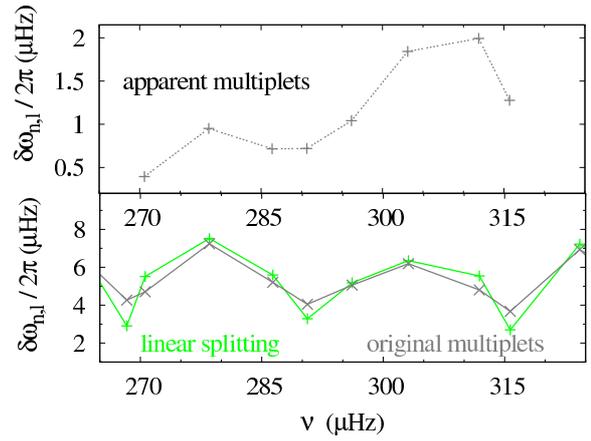}
}  
  \caption{ For the model M$_1$ and a central rotation frequency of $20 \rm \mu$Hz, {\it Top}: spacing between two closest modes of opposite azimuthal order $\rm m = \pm 1$, i.e. {\it apparent} rotational splitting. {\it Bottom}: spacing between prograde and the retrograde modes which belong to the same {\it original} multiplet (grey, crosses) and splittings computed with the first order perturbative method (green, plus signs). 
}
\label{Fig_3}
\end{figure}

Between 8 and 20 $\rm \mu$Hz (Fig. \ref{Fig_suivi_rot}), the mode frequencies seem to behave linearly with $\Omega_c$, but are no longer gathered by {\it original} multiplet. 
Figure \ref{Fig_3} (top) displays the {\it apparent} rotational splittings, taking half the difference between the closest modes of opposite azimuthal order $m=\pm 1$. The values taken by the {\it apparent} rotational splittings range from few hundreds nHz to 2 $\rm \mu$Hz. This is of the order of the splittings measured in observed spectra. Because the apparent multiplets occur by `accident', the curve in Fig. \ref{Fig_3} (top) does not --and usually cannot-- follow the particular `V' pattern around the p-m modes as observed for some slowly rotating red giants (see for observations \citealt{Beck2012}, \citealt{Mosser2012a}). This `V' pattern has the same origin as the modulation seen for mode inertia (\citealt{Goupil2013}) that is due to the trapping of modes \citep[see][]{Dziembowski2001,Dupret2009}. The `V' pattern can then constitute a strong indication that the selection of multiplet has been done correctly in observed spectra. Figure \ref{Ec_trip} confirms that modes  which belong to the same {\it original} multiplet (circled in black) are found in different {\it apparent} multiplets, and show quite different p-g nature. Fig \ref{Fig_3} (bottom) compares the {\it original} splittings --i.e. half the differences between frequencies of modes from the same {\it original} triplet with opposite $m$-- with the splittings given by the first order approach. The g-m modes (e.g. at 277 $\mu$Hz and 303 $\mu$Hz)  have the largest splittings. For such modes, both computations give very close results (with a discrepancy of $\sim 4 \%$). The p-m modes (e.g. at 290 $\mu$Hz and 315 $\mu$Hz) have the smallest splittings and show the expected `V' shape variation. For these modes, the discrepancies can reach $30 \%$. This is due to the trapping of these modes which differs from one $m$ to the other. Note that {\it original} splittings including three terms ($M=3$) in the non-perturbative calculations give relative differences lower than $10^{-6}$ when compared with results of computations using two $Y_{\ell}^m$ (Fig. \ref{Fig_M1-M3}).

Therefore, for a rotation between 8 and 20 $\rm \mu$Hz, the fact that the multiplets are correctly selected can be assumed if the splitting given by {\it apparent} triplets do follow the `V' pattern. If so, the first order approach provides an order of magnitude estimate of the core rotation rate, but non-perturbative modeling is required for accurate quantitative conclusions on the rotational profile from p-m modes.

\section{Rapid core rotating red giants}

\begin{figure}[t!]
{\includegraphics[scale=0.85, angle=-0]{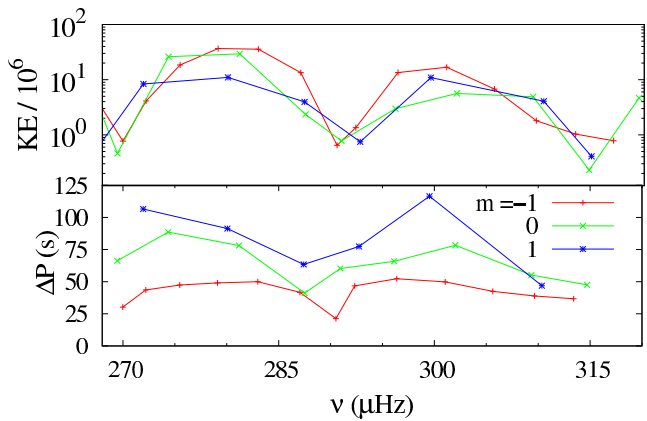}
}  
  \caption{For a central rotation frequency of 140 $\rm \mu$Hz, {\it top}: pulsation modes' kinetic energy with respect to their frequency. {\it Bottom}: period spacing with respect to the pulsation frequency.}
\label{Fig_5}
\end{figure}

For $\Omega > 20 \mu$Hz, the mode frequencies no longer behave linearly with the rotation rate (see Fig. \ref{Fig_suivi_rot}). This is due not only to higher order effects that come into play, but also to the trapping of modes that is modified by rotation. Modes which belong to the same {\it original} multiplet do not have the same radial distribution. Under these circumstances, the notion of rotational splitting  as defined by $\delta \omega_{n,\ell}^1$ is no longer relevant, and cannot be simply related to the rotation profile.
Figure \ref{Fig_5} (top) shows the kinetic energy of pulsating modes for a core rotation rate of 140 $\rm \mu$Hz. The large separation, as measured by the difference between consecutive p-m modes with minimum kinetic energies, is conserved. The higher kinetic energy of prograde modes indicates that they are more of g nature than retrograde or axisymmetric ones. This is due to the shift in frequency induced by rotation that brings, in the same frequency range, modes which belong to very different parts of the zero rotation spectrum (see Fig. \ref{Fig_suivi_rot}).

Figure \ref{Fig_5} (top) shows that, in the same range of frequency, there are more $m=-1$ modes than $m=0$ modes, and in turn, there are more $m=0$ modes than $m=+1$ ones. In order to highlight this difference in distribution, the period spacings are plotted for each $m$ value in Fig. \ref{Fig_5} (bottom). 
In this diagram, the families of modes of different azimuthal orders clearly show different values of period spacing, ranging from the lowest for the prograde modes to the highest for the retrograde ones. Note that this phenomenon appears only for a rotation high enough that the distributions of the three $m$-families of modes become clearly different ($\Omega_c \sim 100 \mu$Hz for  model M$_1$). If one is able to measure three different values of period spacings in a observed spectrum, one is able to identify the values of the azimuthal order. Moreover, based on our study of synthetic spectra, we find that the differences between the values of period spacings increase with $\Omega_c$, and therefore can constitute a guideline toward a constraint on the rotation profile.

\section{Discussions and conclusions}
\label{S4}

Some observed  spectra of red giant stars show structures with nearly symmetric spacings around axisymmetric modes. For this kind of spectra, we found that one can figure out if these  {\it apparent} multiplets correspond to  the {\it original} splittings by plotting the {\it apparent} splittings as a function of the axisymmetric modes frequencies. If the resulting curve follows the `V' pattern, then it is reasonnable to assume that the triplets are correctly selected, and quantitative information on rotation can be derived from  the observed splittings. This corresponds to a low rotation regime when the splitting is smaller than the frequency spacing between two consecutive modes of same $m$, i.e. for ${\rm \Omega_c/(2 \pi) / (\Delta P / P^2)} \lesssim 2$.

 If the {\it apparent} splittings do not follow the `V' pattern, the triplets overlap. The star is therefore located in the regime of moderate rotation. The frequency differences between the {\it original} triplet members are large enough that the trapping properties differ from one member to another. However, if it is still possible to select correctly the {\it original} triplets \citep[as done in][]{Mosser2012b}, the first order approach could be used for a first guess of the core rotation rate. Non-perturbative calculations are nevertheless required for a precise determination, particularly because of the avoided crossings that p-m modes undergo. This rotation regime corresponds to values ${\rm \Omega_c/(2 \pi ) / (\Delta P / P^2)}$ between 2 and 5. 

Finally, some observed spectra of red giants do not show any regular or close to regular structures. We expect these spectra to correspond to rapid rotators. In this case, modes of different azimuthal orders have a very different nature. The concept of rotational splitting of modes with the same radial distribution and different $m$ is no longer relevant, and one should consider separately sub-spectra associated to each value of $m$. Provided that the rotation is high enough to give rise to clearly different distributions with respect to $m$, the differences between the period spacings associated with $m$ allows to identify the azimuthal order, and thereby offers the promising opportunity of deriving a proper seismic diagnostic. That corresponds to the very rapid rotation case i.e. for ${\rm \Omega_c/(2 \pi ) / (\Delta P / P^2) \gtrsim 20}$. An intermediate case remains, where the rotational splitting is no longer a relevant seismic diagnostic, but where the rotation is too slow to allow to distinguish three different period spacings associated to the $m$-families of modes (for ${\rm \Omega_c/(2 \pi) / (\Delta P / P^2)}$ between 5 and 20).  Establishing a diagnostic in this regime constitutes an important issue  which needs to be adressed in a forthcoming paper.

It can be surprising that when $ \Omega/ (2\pi) / (\Delta P / P^2) < 20$, on the one hand the perturbative approach gives inaccurate results, but on the other hand non-perturbative models where the spectral expansions are limited to one term (plus the toroidal component) give accurate enough results. This can be understood with the help of the following equation: $\Omega(r)/2\pi \sim \Delta \nu \ll \nu_{max}$. In the  frequency range of interest here, the rotation frequency remains significantly lower than $\nu_{max}$ throughout the whole star, and thus lower than the pulsation frequencies. Hence, the Coriolis terms are smaller than other terms in the equation of motion and the coupling of $\ell=1$ modes with $\ell=3,5,...$ components is small. 
However the rotational splitting is not small compared to the large separation $\Delta \nu$. This implies that the trapping of each member of a same triplet can be very different, as illustrated in Fig.\ref{Ec_trip}. 
Using the variational principle to model rotational splittings is thus not justified because the eigenfunctions are too different. The only way to properly model the effect of rotation on oscillations is to solve separately the differential equations associated to each member of the multiplet.

 To sum up, the major impact of moderate rotation on red giants' spectra is the modification of the trapping that depends on $m$. Let us precise that here the cavities are not considered to be modified by rotation, but it is the way modes probe them which differs from one member of a multiplet to another. In such a case, the first order perturbative approach gives inaccurate results and the $m$-families of modes carry different information on the stellar interior. Only methods which compute these $m$-sub-spectra independently, i.e. non-perturbative methods, are appropriate for studing the moderate to rapidly core rotating red giants.

\vspace*{-0.2cm}
\begin{acknowledgements}
RMO thanks Beno\^it Mosser for fruitful discussions. 
The authors thank the referee for his comments that helped to improve the manuscript.
\end{acknowledgements}
\vspace*{-0.2cm}

\bibliographystyle{aa} 
\vspace*{-0.5cm}
\bibliography{rmomj-splitting_RG-AAmain_v18_arxiv} 

\end{document}